\documentclass[reprint,amsmath,amssymb,aps,prb]{revtex4-2}

\usepackage[utf8]{inputenc}
\usepackage[T1]{fontenc}
\usepackage{graphicx}

\usepackage{times}
\usepackage{mhchem}[version=4]
\usepackage{hyperref}
\usepackage{booktabs}
\usepackage{xcolor}
\usepackage{newunicodechar}
\newunicodechar{–}{--}
\begin{document}

\title{Lattice and Orbital-Resolved Fermiology of Metallenes}

\author{Kameyab Raza Abidi$^1$}
\email{kameyab.r.abidi@jyu.fi}
\author{Mohammad Bagheri$^1$}
\author{Pekka Koskinen$^1$}
\email{pekka.j.koskinen@jyu.fi}
\affiliation{
$^1$Nanoscience Center, Department of Physics, University of Jyväskylä, Finland.
}

\begin{abstract}

Atomically thin metallenes have emerged as a new member of the two-dimensional (2D) materials family. Recent experimental realization of metallenes in the Ångström limit has further intensified interest in this class of 2D materials. However, achieving sub-atomic insight into them demands the most detailed and systematic characterization of their electronic structure. Such understanding is essential for the rational design and exploitation of their properties in plasmonics, catalysis, and quantum optics.
Existing electronic-structure studies are either scattered or focus on a few selected systems, and a comprehensive view of their band structures and Fermi surfaces 
remains missing. Here, we address this gap by studying 45 elemental metallenes in six monolayer lattices (honeycomb, square, hexagonal, and their buckled forms) using density-functional theory. 
We found that lattice type primarily fixes the shape and radial placement of the Fermi-lines, while out-of-plane buckling introduces controlled modifications: it shortens long straight Fermi-line segments, and occasionally creates, removes, or merges small Fermi-line pockets. The electronic configuration determines which orbital type dominates the Fermi level.  
We summarized Fermiology using a single score for each element, termed pocketness, derived from four descriptors that combine element properties (symmetry, coordination) with electronic characteristics (dispersion, Fermi-surface topology). This score enables targeted angle-resolved photoemission spectroscopy (ARPES) tests, controlled Lifshitz transitions, and provides a predictive basis for transport and device applications.

\end{abstract}

\date{\today}

\maketitle

\section{Introduction}

In recent years, tens of elemental metallenes have been realized as single–atom layers, both free-standing (within protective pores or after etching) and substrate-supported—marking an experimental milestone~\cite{metallenes_rev_AM_2025, 2D_metals_2025, kashiwaya2024synthesis}. Demonstrations span light $s$ metals (Be, K), transition metals (Cr, Fe, Zr, Mo, Rh, Hf, Os, Au), and heavy $p$–block elements (In, Sn, Tl, Pb, Bi), with representative syntheses via \emph{in situ} e-beam sculpting inside graphene pores (Cr, Fe, Sn, Zr, Mo), epitaxial growth of ordered monolayers on single-crystal supports [Hf on Ir(111); Au and Ga; superconducting In/Pb on Si(111); bismuthene on SiC], wet-chemical routes to single-layer metals (goldene; monolayer Rh), self-organized monolayers on covalent two-dimensional (2D) hosts (Os on graphdiyne), and controlled intercalation producing metallene mono or multi layer alkalis and related films (K, Li, Mn, Bi, Na) \cite{ta2020Cr,zhao2014Fe,ta2021situ,yang2021Sn,mendes2024Zr,zhao2018Mo,li2013Hf,preobrajenski2024boron-Au,tao2018gallenene,zhang2010superconductivityPb-In,reis2017bismuthene,kashiwaya2024synthesis,duan2014Rh,gao2024selfOsmium,yin2015real,yin2015simple,chahal2023beryllene, pathirage2025-Mn-intercalation, 2D_metals_2025, Interc_Na_trilayer_2025}.

Elemental monolayers of metals offer potential for transparent conductors, plasmonic sheets, electrocatalysts, quantum optics, and spin–orbit platforms \cite{zhang2010superconductivityPb-In,reis2017bismuthene, syperek2022observation,ruckman1994monolayer}. All these features are mainly controlled by electronic bands which set the Fermi-lines and velocity $v_F$, effective masses, nesting vectors, and the neighborhoods where hybridization opens mini-gaps or flattens dispersion~\cite{Dugdale_2016,FS_nesting_2001,whangbo1991hidden,zhu2015cdw, terashima2009fermi}. Previous experimental and theoretical work has examined the electronic structure of individual systems~\cite{abidi2023electronic,li2014electronic}~e.g., Hf on Ir(111) by ARPES~\cite{Xiao2021_Hf_on_Ir(111)}, gallenene on specific substrates~\cite{tao2018gallenene,li2019ultrathin-Ga, petrov2021superconductivity}, heavy sheets $p$~\cite{FS_Pb_2025,FS_Stanene_2018,terakawa2018identification_Indium}, ultrathin noble metals~\cite{ono2025framework_FS_Au_Al, conductivity_goldene_2024, kashiwaya2024synthesis,FS_Pt_2021,preobrajenski2024boron-Au,abidi2025electronic}. However, the field still lacks a systematic, comprehensive study across many metallenes and multiple lattices.

Here, we set out to establish such a baseline by identifying the recurring features of bands and Fermi-lines across standard lattices, i.e., honeycomb (hc), square (sq), hexagonal (hex), and their buckled forms (Figure~\ref{fig:bands_fermi_schematics}) as well as across $s$-, $p$-, and $d$-led metallenes.

A central aim is to quantify features such as band crossings at the Fermi level~($E_F$), band flatness, anisotropy, and $\Gamma$-centricity in a manner that remains comparable across different elements and lattice types.

Establishing these quantities also enables a rational selection of metallenes exhibiting targeted electronic characteristics, whether for achieving nearly isotropic transport or for accessing regimes where quantum effects such as Shubnikov–de Haas oscillations \cite{bangura2008small} become essential.

Therefore, in this article, we employed density-functional theory simulations to study monolayers of 45 elements (Groups 1–15) in six standard lattices, resulting in 270 monolayers. We calculated orbital-resolved bands and Fermi-line maps with Fermi velocity $v_F$, to observe trends across chemistries and symmetries. We found that element dictates which orbitals have states at $E_F$:~$s/p$- for Groups 1–2 and 11, $d$- for the transition series, and $p$ for Groups 13–15. The lattice type dictates where the main Fermi-line features sit and how stretched they are: hex tends to put more weight near $\Gamma$-point and shows the most band crossings across $E_F$, sq tends to form long, edge-aligned segments, and hc yields the most compact Fermi-line topology, typically a single loop with minimal additional Fermi-lines. Buckling introduces modest changes and reshapes bands and Fermi-lines: it shortens the edge-aligned Fermi-line segments, slightly raises local band flatness, and sometimes creates or merges small Fermi pockets without altering the in-plane symmetries. These trends yield practical guidelines for selecting metallenes and lattices, and the results will serve as the general descriptors for band structure and Fermi-lines of metallenes.

\section{Computational Methods}

All calculations were performed within density-functional theory (DFT) using QuantumATK~(v.~2023.12)~\cite{smidstrup2019quantumatk}. The exchange–correlation energy was approximated by Perdew–Burke–Ernzerhof generalized gradient approximation (GGA–PBE)~\cite{perdew1996generalized, zhang1998comment} for geometry optimization, and the Heyd–Scuseria–Ernzerhof hybrid functional (HSE06)~\cite{HSE06} for the bands and Fermi contours calculations.~Side-by-side PBE and HSE06 band-structure plots provided in the Supplemental Material~\cite{SM} show that the qualitative low-energy features underlying our analyses are consistent, while HSE06 primarily refines relative band positions and splittings.
Core–valence interactions were described by \textsc{PseudoDojo} norm–conserving pseudopotentials including scalar–relativistic effects~\cite{van2018pseudodojo}. The single–particle states were expanded in a linear combination of atomic orbitals (LCAO) basis using the \emph{Medium} set supplied by QuantumATK. Brillouin–zone sampling employed a $13\times13\times1$ Monkhorst–Pack grid~\cite{Monkhorst_Pack} with Fermi–Dirac smearing of $0.05$~eV  for SCF calculations and a dense uniform $k$-grid of $101 \times 101 \times 1$ for Fermi-line mapping in the FBZ.

Each monolayer was modeled with $20$-\AA\ vacuum normal to the monolayer plane. Atomic positions and in–plane lattice vectors were relaxed until the maximum residual force was below $10^{-6}$~eV/\AA\ and the residual in–plane stress below $10^{-3}$~eV/\AA$^{2}$. 
We adopted the optimal DFT parameters established in earlier work~\cite{abidi2022optimizing}.
Band structures were computed along standard two-dimensional high–symmetry paths~(Figure~\ref{fig:bands_fermi_schematics}). 

The energetic and dynamical stabilities of atomically thin metallenes have already been investigated systematically~\cite{nevalaita2018atlas, ono2020dynamical}. Extensive energetic and dynamical stability analyses for all 270 monolayers are presented in Ref.~\cite{abidi2024gentle}, which shows that almost half of these lattices are dynamically stable as freestanding membranes, either strained or unstrained. For completeness, here we analyze the electronic structures of all 270 unstrained lattices, including the dynamically unstable ones. We adopt this inclusive approach because metallenes are frequently stabilized by extrinsic factors such as substrate interactions, encapsulation, or confinement within pores, which can suppress the phonon instabilities predicted for ideal membranes~\cite{2D_metals_2025, metallenes_rev_AM_2025, kashiwaya2024synthesis, li2013Hf, ta2021situ}. 
An illustrative example is of thallene, a graphene-like honeycomb lattice of Tl reported to be stabilized on a NiSi$_2$ substrate~\cite{gruznev2020thallene}, while the corresponding freestanding honeycomb Tl monolayer is dynamically unstable as predicted by phonon bands even when strained~\cite{abidi2024gentle,ono2020dynamical}. Therefore, capturing stabilization mechanisms from specific external perturbations requires case-by-case modeling (substrate/encapsulation/environment).

Consequently, given the realization through diverse experimental synthesis strategies, the band structures and Fermi surfaces presented herein serve as a baseline for systematic understanding of the electronic properties of these systems.

\subsection{Fermiology metrics from bands and Fermi-lines}
\begin{figure*}[t!]
      \centering
    \includegraphics[width=0.99\linewidth]{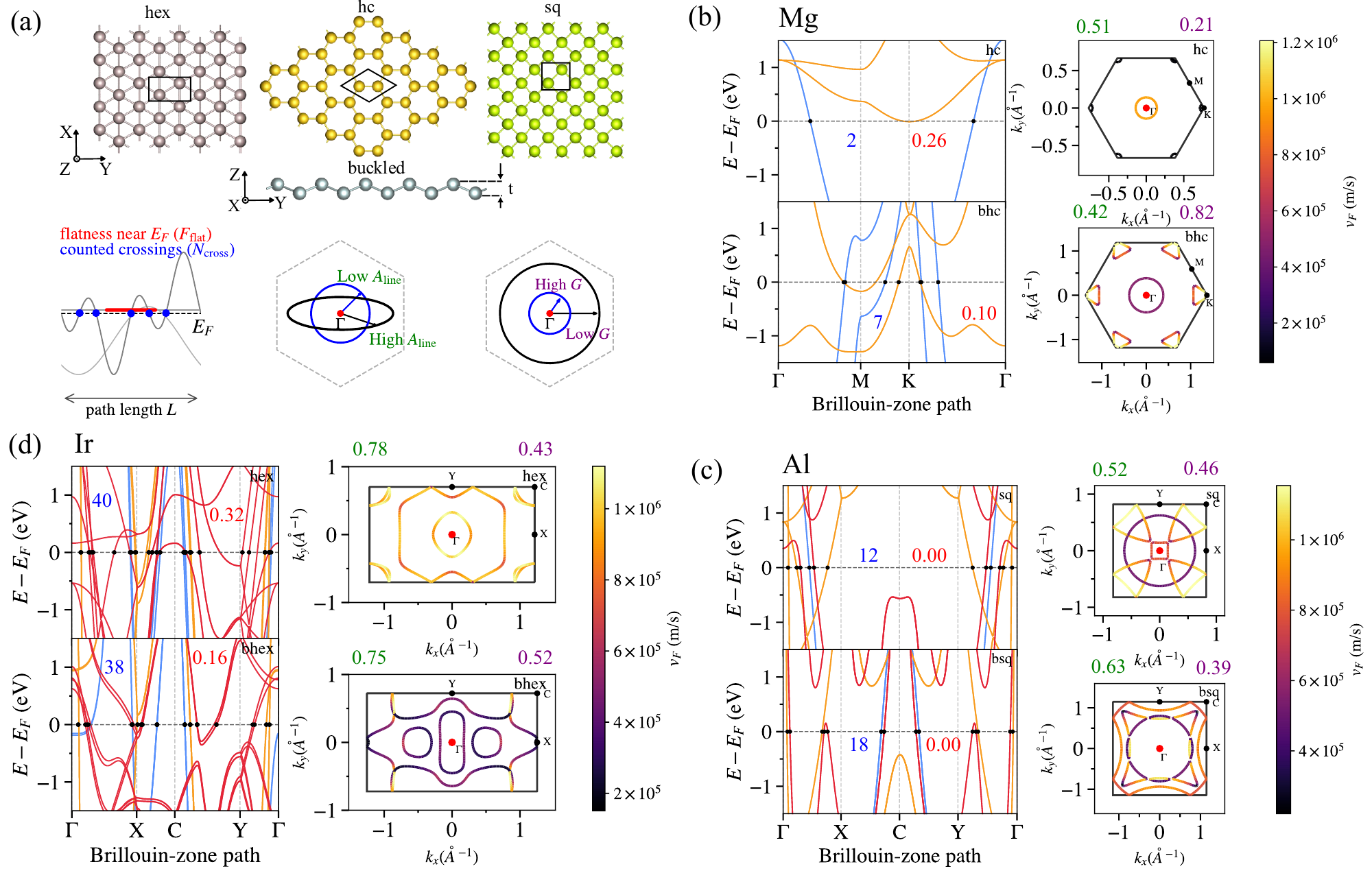}
    \caption{(a)~Structural and Fermiology metric schematics. Top: hexagonal (hex), honeycomb (hc) and square (sq) metallenes with computational unit cells (black quadrilateral) and a side view illustrating out-of-plane buckling. Bottom: schematics defining the four fermiology metrics: $N_{\mathrm{cross}}/L$ (number of band crossings at $E_F$ per path length), $F_{\mathrm{flat}}$ (fraction of the near-$E_F$ path spent on flat segments), $A_{\mathrm{line}}$ (degree of Fermi-line isotropy, low for nearly isotropic ($\Gamma$-centred loops, high for edge-aligned segments), and $G$ ($\Gamma$-centricity, i.e. weight of Fermi lines near $\Gamma$ versus the zone edge). The grey dashed hexagon schematically marks the first Brillouin zone. ~Representative band structures and Fermiologies of metallenes.~(b) Mg in honeycomb (hc) and buckled honeycomb (bhc) lattices, (c) Al in square (sq) and buckled square (bsq) lattices, and (d) Ir in hexagonal (hex) and buckled hexagonal (bhex) lattices. For each case, the left panel shows the band structure along the high-symmetry path with the Fermi level fixed at $E_F = 0$; bands are colored by orbital type projection as implemented in QuantumATK~\cite{smidstrup2019quantumatk} (blue ($s$), orange ($p$), red ($d$)).~The right panel shows the corresponding Fermi-line map in the first Brillouin zone~(FBZ) at $E_F$, here the color encodes the Fermi velocity ($v_F$). The red dot marks the $\Gamma$-point at the FBZ center, and black dots on the FBZ boundary indicate high-symmetry points: edge centers are $M$ for hc/bhc and $X/Y$ for sq/bsq/hex/bhex, while zone corners are $K$ for hc/bhc and $C$ for sq/bsq/hex/bhex.
For each band-structure and Fermi-line plot, the values of $N_{\mathrm{cross}}/L$~(blue) and $F_{\mathrm{flat}}$~(red), $A_{\mathrm{line}}$~(green) and $G$~(purple) are shown. The complete set of bands and Fermi maps for all metallenes and lattices is in the SI.
}
    \label{fig:bands_fermi_schematics} 
\end{figure*}
We map the Fermi-lines in the first Brillouin zone (FBZ) by solving the equation $\varepsilon_n(\mathbf{k})=E_F$. The band velocities are obtained from first-order perturbation theory for the generalized eigenproblem
\begin{equation}
H(\mathbf{k})\,\mathbf{c}_n(\mathbf{k})=\varepsilon_n(\mathbf{k})\,S(\mathbf{k})\,\mathbf{c}_n(\mathbf{k}),
\end{equation}
where $H(\mathbf{k})$ is the Bloch Hamiltonian in the chosen basis, $S(\mathbf{k})$ is the overlap matrix, $\mathbf{c}_n(\mathbf{k})$ the eigenvector of expansion coefficients for band $n$, and $\varepsilon_n(\mathbf{k})$ the single-particle eigenenergy. The band velocity components are

\begin{equation}
v^{\alpha}_{n}(\mathbf{k})
=\frac{1}{\hbar}\,
\big\langle \psi_{n}(\mathbf{k}) \,\big|\, \partial_{k_{\alpha}} H(\mathbf{k}) - \varepsilon_{n}(\mathbf{k})\, \partial_{k_{\alpha}} S(\mathbf{k}) \,\big|\, \psi_{n}(\mathbf{k}) \big\rangle,
\end{equation}
with \(\alpha\in\{x,y\}\) for 2D, and $|\psi_n(\mathbf{k})\rangle$ the normalized Bloch state associated with $\mathbf{c}_n(\mathbf{k})$. The Fermi velocity at a point $(n,\mathbf{k})$ on the Fermi contour is the Euclidean norm of the velocity vector,

\begin{equation} v_F(n,\mathbf{k})=\big\|v_n(\mathbf{k})\big\|=\sqrt{\sum_{\alpha}\big(v_n^{\alpha}(\mathbf{k})\big)^2}.
\end{equation}
For brevity, we write $v_F$ with the band/index $(n,\mathbf{k})$ implicit.

For each element–lattice system $(Z,l)$ we compute the average Fermi velocity
\begin{equation}
\langle v_F\rangle_{Z,l}
=\frac{1}{|\mathcal{F}_{Z,\ell}|}\sum_{(n_i,\mathbf{k}_i)\in \mathcal{F}_{Z,\ell}} v_F(n_i,\mathbf{k}_i),
\end{equation}
where $\mathcal{F}_{Z,\ell}$ is the number of sampled Fermi-line points collected for element $Z$ in lattice $\ell$.

Further, to quantify how anisotropic the Fermi-lines are, we introduce the Fermi-line anisotropy $A_{\mathrm{line}}(Z,\ell)$. For a given element-lattice system, each connected Fermi-line segment at $E_F$ is analysed to determine whether its shape is (nearly) circular or strongly elongated along a preferred direction. For each segment, we took all sampled $\mathbf{k}$-points, normalized them by a characteristic Brillouin-zone length scale (the maximum $|\mathbf{k}|$ within the FBZ) so that different lattices are comparable, and then evaluated how broadly these points are distributed along the in-plane $k_x$ and $k_y$ directions. We condensed this geometric information into a single anisotropy measure $A_{\mathrm{line}}$~(details are in SI). Low values of $A_{\mathrm{line}}$ correspond to nearly isotropic, loop-like segments, whereas high values indicate strongly elongated, almost line-like segments~(Figure~\ref{fig:bands_fermi_schematics}a).  For example, V(hc) or coinage metals (Cu, Ag, and Au) in bsq have almost isotropic $\Gamma$-centred loops with very low $A_{\mathrm{line}}$. In contrast, Cu(hc), Fe(hc), and Co(hc) exhibit Fermi-line segments that are confined to narrow regions near the Brillouin-zone edges and corners; the spectral weight around the zone centre is anisotropic. These cases correspond to the large values of $A_{\mathrm{line}}$. It is worth mentioning that, by construction, $A_{\mathrm{line}}$ does not change if the whole Fermi surface is shifted or rotated in $\mathbf{k}$-space, and it is independent of the absolute size of the Brillouin zone.

To characterize where the Fermi-lines are positioned within the FBZ, we introduce the $\Gamma$-centricity $G(Z,\ell)$. We took all sampled $\mathbf{k}$-points for a given Fermi-line segment and measured their distance from the zone centre $\Gamma$. These distances are expressed in units of a characteristic Brillouin-zone length scale, given by the largest $|\mathbf{k}|$ on the zone boundary, so that all lattices share a common radial scale. We then summarize this set of radii by a single value and convert it into a metric that is large when most Fermi-line segments lie close to $\Gamma$ and small when they are concentrated near the zone edges and corners~(Figure~\ref{fig:bands_fermi_schematics}a). In other words, high $G$ indicates segments near $\Gamma$-point, whereas low $G$ means Fermi-lines live mainly in the outer part of the zone or near the edges. Because the construction only depends on distances from $\Gamma$, $G$ does not change if the Fermi surface is rigidly rotated in $\mathbf{k}$-space~(details are in SI). 
For example, Sr and V in hc have large $G~\sim 0.80$, showing that their Fermi-lines are around $\Gamma$-point with a very small weight at the boundary. By contrast, Mo, W, and Cu in hc have small $G\approx 0.01-0.08$, in these cases the Fermi-lines are almost entirely confined to narrow segments near the Brillouin-zone edges and corners, with very little spectral weight in the central region. Between these extremes lie systems such as Zn(hex) and Cd(hex), where $G\approx 0.54-0.59$, indicating Fermi-line segments sit well inside the zone boundary but are not tightly localized at $\Gamma$.

For quantifying the complex arrangement of dispersive bands in the vicinity of the Fermi level, we defined the crossing density for an element-lattice pair as \(N_{\mathrm{cross}}(Z,\ell)/L\).
The quantity \(N_{\mathrm{cross}}(Z,\ell)\) counts how many times the band crosses $E_F$ along the chosen high-symmetry path, and then normalizes this count by the total path length $L$ so that different lattices can be compared directly. In practice, we examined each band along the path and recorded every instance where its energy switches from below to above the Fermi level (or vice versa), excluding cases which fall within a very small numerical window around $E_F$ to avoid noise~(mathematical details are in SI). A larger value of \(N_{\mathrm{cross}}/L\) indicates an intricated low-energy band structure with many Fermi-level intersections~(Figure~\ref{fig:bands_fermi_schematics}). For example, Bi(hc) has very low crossing density with \(N_{\mathrm{cross}}/L~=~2\) and for Os(hex)/Y(sq) crossing density is 50/44, illustrating how \(N_{\mathrm{cross}}/L\) distinguishes between sparse and dense multiband cases near Fermi-level.

Finally, we quantify how flat the bands are near the Fermi level. Flat bands enhance the density of states and provide slow carriers, so we want a single number that tells us what fraction of bands the near $E_F$ path is spent on flat pieces rather than steep, dispersive ones. For each element $Z$ in lattice $\ell$, we follow the bands along the high-symmetry path and keep only those points whose energy lies within a fixed window of $E_F~\pm~0.30\ ~\text{eV}$. At every Fermi-level crossing, we remove a small neighborhood so that sharp kinks at the crossings do not artificially count as “flat”. On the remaining portions of the path, we evaluate the band slope along the path and label as “flat” those stretches where the slope stays below a threshold, chosen from the typical near $E_F$ slopes in that system. Isolated points are discarded; only segments that remain low-slope over several consecutive $k$-points are kept. The flatness metric $F_{\mathrm{flat}}$ is then defined as the fraction of the near $E_F$ path length that lies on such flat stretches, averaged over the continuous segments of the path~(details are in SI).

Values of $F_{\mathrm{flat}}$ close to zero mean that the bands in the range $E_F \pm 0.30\ \text{eV}$ are uniformly steep and dispersive; for example, Al(sq/bsq) and Na(hc/sq/hex) have essentially zero $F_{\mathrm{flat}}$~(Figure \ref{fig:bands_fermi_schematics}c).  Intermediate values reflect a mixture of steep and flat pieces; V(sq) and Re(hex) fall into this regime with $F_{\mathrm{flat}}\approx 0.26$ and $0.27$, respectively.  The large value occurs when extended low-slope plateaus run along the path, as in Mo(sq), where $F_{\mathrm{flat}}\approx 0.52$.  In this way $F_{\mathrm{flat}}$ complements $N_{\mathrm{cross}}/L$: the latter counts how many bands reach the Fermi level, while $F_{\mathrm{flat}}$ measures how much of that near $E_F$ spectrum is flat.

\section{Results}

We now present a systematic analysis of the electronic structures of 45 elemental monolayers placed in six two-dimensional lattices: planar honeycomb (hc), square (sq), hexagonal (hex), and their buckled counterparts. 
For each system, we examine orbital-resolved band structures and Fermi-lines with Fermi-velocity colormaps~(Figure~\ref{fig:bands_fermi_schematics}, and more details in SI). The discussion is organized by the \emph{orbital identity} of the states at $E_F$. This groups the 270 monolayers into five thematic groups:
(i)~nearly free-electron $s$-metallenes (Groups~1--2),
(ii)~closed-$d$-shell metallenes (Groups~11--12),
(iii)~$p$-band metallenes~(Groups~13--15),
(iv)~early transition $d$-metallenes (Groups~3--6), and
(v)~late transition $d$-metallenes (Groups~7--10).
Representative band structures and Fermi-line maps for selected element–lattice systems are shown in Figure~\ref{fig:bands_fermi_schematics}(b–d) and a complete set in the SI, while Figures~\ref{fig:vf_lattice_orbital}–\ref{fig:pocketness} summarize the full data set as family-resolved $\langle v_F\rangle$, lattice- and orbital-resolved medians of the four metrics, their mutual correlations, and the periodic-table pocketness map. 
We start with the simplest single-manifold baseline to the most intricate multi-sheet $d$-band systems, keeping coordination and orbital hierarchy in view. We begin with the nearly free-electron $s$- metallenes. 
\subsection{$s$-metallenes (Groups 1--2)}
\begin{figure}[t!]
      \centering
    \includegraphics[width=0.868\linewidth]{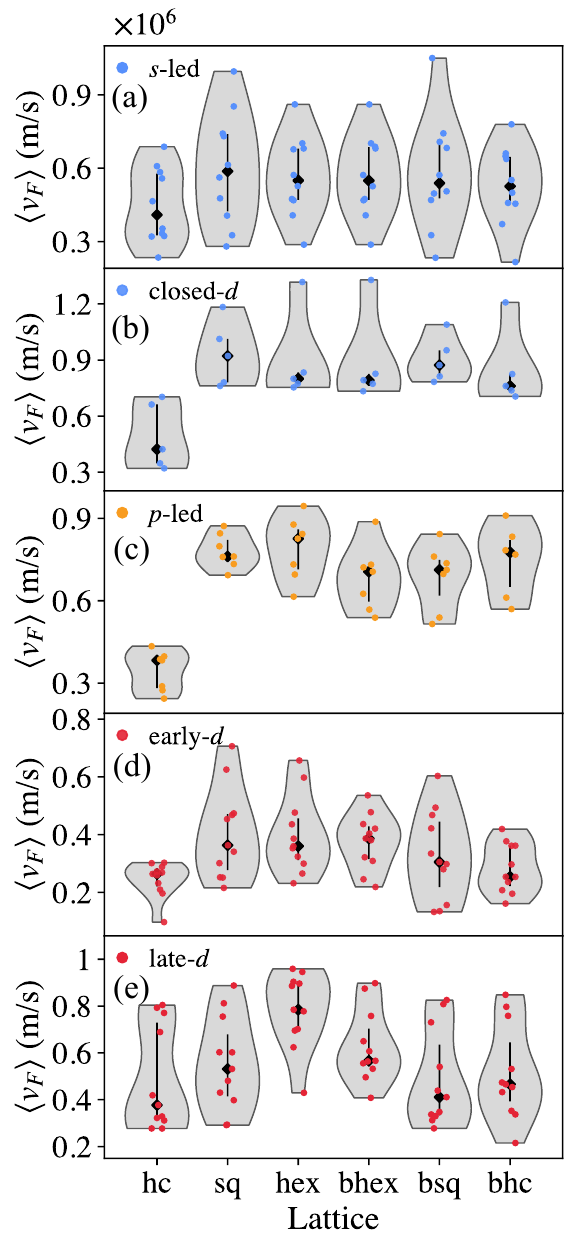}
    \caption{~Violin plots show the distribution of $\langle v_F\rangle$ (m/s) for elemental metallenes grouped by valence family. (a) $s$-led~(Groups 1–2),
(b) closed-$d$~(Groups 11–12),
(c) $p$-led~(Groups 13–15),
(d) early-$d$~(Groups 3–6), and
(e) late-$d$~(Groups 7–10).
In each panel, the six lattices (hc, sq, hex, bhex, bsq, bhc) appear on the horizontal axis. Individual symbols indicate single metallenes, the central diamond marks the median, and the vertical bar shows the interquartile range for that lattice and family.~}
    \label{fig:vf_lattice_orbital} 
\end{figure}

Groups 1–2 show a $\Gamma$-centered, dispersive $s$-band that sets the low-energy structure, while $p$- and, in heavier alkaline earths, near-$E_F$ $d$-state channels enter only where symmetry allows; the Fermi contour reflects this backbone as either a single $\Gamma$-loop or a loop plus small pockets. The average Fermi velocities $\langle v_F\rangle$ are relatively high and only weakly lattice-dependent: for most $s$-metallenes they fall in the range $(3$--$6)\times10^{5}~\mathrm{m/s}$ across lattices, with slightly higher medians in sq/hex/bhex than in hc/bhc ~(Figure~\ref{fig:vf_lattice_orbital}a).

In the alkalis (Li–Cs), the $F_{\mathrm{flat}}$ stays small across lattices (mostly $\approx 0$), with a modest rise in bhc and bsq for Li/Rb/Cs, which corresponds to the uniformly dispersive $s$-branch. The Fermi-level crossing density $N_{\mathrm{cross}}/L$ stays low in Group 1. In the hc lattice for Na–Cs, the bands meet $E_F$ only at the zone-corner $K$, so $N_{\mathrm{cross}}/L~=~1$ and the Fermi contours disappear; no visible Fermi-line segment exists.
Sq/hex/bhex in turns give a single $\Gamma$-loop with $N_{\mathrm{cross}}/L~\approx~2-4$. Fermi-line shape is captured by $A_{\mathrm{line}}$: it is very high in sq and hex/bhex for the alkalis (high elongation set by lattice warping) but becomes nearly circular loops in bhc, which reflects the more rounded concentric sheets seen after buckling~(Figures.~\ref{fig:bands_fermi_schematics}b,~\ref{fig:metrics_lattice_orbital}a,   and ~S1-S5). The calculated $\Gamma$-centricity, $G$, shows that the loops are toward the outer half of the FBZ in sq and hex/bhex and shifts them slightly inward in bsq; in hc, Li shows edge-grazing Fermi-line arcs (small $G$). Overall, the alkali metallenes have low $F_{\mathrm{flat}}$ and $N_{\mathrm{cross}}/L$, indicating only a few dispersive bands with few Fermi-level crossings. Calculated $A_{\mathrm{line}}$ is high in sq/hex lattices, meaning the presence of an elongated $\Gamma$-centered loop, and low $A_{\mathrm{line}}$ in bhc implying a near-circular loop. Small-to-moderate $G$ means that the Fermi-lines lie in the outer half of the FBZ. Buckling changes loop geometry and modifies a single $\Gamma$-centered loop into two concentric loops, raising $N_{\mathrm{cross}}/L$, however, the near-$E_F$ states remain $s$-dominated ~(Figure~\ref{fig:bands_fermi_schematics}b and Figures~S1-S5).

Alkaline earths metallenes (Be–Ba) move to a multiband regime as coordination and buckling bring $p$- and $d$-bands into the $E_F$ (Figures S6-S10~\cite{SM}). $F_{\mathrm{flat}}$ rises from near zero up to ($\sim 0.26$) in several lattices, consistent with local slope suppression near avoided crossings, which is visible in the band plots. The $N_{\mathrm{cross}}/L$ increases markedly, and is often highest in bsq and bhc~(Figure~\ref{fig:metrics_lattice_orbital}a). Relative to the alkalis, the Fermi-line geometry of alkali metallenes differs: $A_{\mathrm{line}}$ becomes lattice-dependent (rounder in hex/bhex when a central loop dominates, more elongated in bsq when edge lenses and corner petals appear~~(Figure~\ref{fig:metrics_lattice_orbital}a), and values of $G$ indicate that contours move inward. These changes follow what the Fermi-contour plots show as added “side loops” and “satellites”: side loops are small closed pockets detached from the main $\Gamma$-loop; satellites are short secondary arcs that closely follow a larger loop. Their appearance raises $N_{\mathrm{cross}}/L$; compact central pockets lower $A_{\mathrm{line}}$ and raise $G$; corner or edge pockets keep $G$ smaller but still lift the crossing count. When buckled, the lattice-paired trends show: breaking mirror symmetry turns the protected contacts at $K$ in hc and $X/Y$ in sq into mini-gaps (former crossings turn into small avoided crossings) and redistributes which branches cross $E_F$, with some crossings moving, appearing, or disappearing. Meanwhile, the Fermi-line topology can switch from a single loop or a point-touching to two loops or pockets. In Group 1 buckling mainly rounds the $\Gamma$-loop (very low $A_{\mathrm{line}}$ in bhc/bsq), while in Group 2 it both rounds or splits loops and adds pockets, so $N_{\mathrm{cross}}/L$ grows and $F_{\mathrm{flat}}$ increases without opening a global gap, as visible in the side-pocket and lens features in the plots (SI).

Thus, Groups 1–2 separate in a four-dimensional descriptor space defined by $F_{\mathrm{flat}},~N_{\mathrm{cross}}/L,~A_{\mathrm{line}}$, and $G$, where alkali metallenes keep small $F_{\mathrm{flat}}$, few crossings, and a single $\Gamma$-centred loop whose elongation depends on lattice and collapses under buckling. Alkaline earth metallenes show larger $F_{\mathrm{flat}}$, higher $N_{\mathrm{cross}}/L$, and lattice-tunable $A_{\mathrm{line}}$ and $G$ that indicate where additional pockets appear.

\subsection{Closed $d$-shell metallenes (Groups 11--12)}

Closed $d$-shell metallenes are mainly governed by $s/p$- bands forming a Fermi loop around $\Gamma$. Compared to the $s$-led metallenes, the closed-$d$-shell metallenes have similar or slightly higher median $\langle v_F\rangle$ for most lattices, consistent with relatively dispersive bands near $E_F$~(Figure~\ref{fig:vf_lattice_orbital}b).

In Cu, Ag, Zn, and Cd the $d$-bands lie far below $E_F$, so they interact with the $s/p$ band only where bands meet or nearly meet at high-symmetry points or lines. In Au, relativistic effects lift $d$-bands toward $E_F$ making $s/p$–$d$ mixing strong.
The Fermi-line contours are either forming a single $\Gamma$-loop or a loop plus small pockets~(Figure~\ref{fig:metrics_lattice_orbital} and Figures S33-S37~\cite{SM}).
For the coinage metals, $F_{\mathrm{flat}}$ is small or vanishing in hex and sq for Cu and Ag, rises in hc where slopes suppress near $K$. Upon buckling, $F_{\mathrm{flat}}$ remains modest, indicating largely dispersive $s/p$-bands. 
The crossing density $N_{\mathrm{cross}}/L$ is high in hex/bhex for Cu ($\approx$~14-18) and lower in sq (12), reflecting the added band intersections allowed by sixfold coordination.

The $A_{\mathrm{line}}$ $= 0.96$ in hex/bhex and $0.98$ in sq (elongated $\Gamma$-loops), but bsq is nearly isotropic with vanishing $A_{\mathrm{line}}$ for Cu/Ag). These loops exist in the outer half of the FBZ in hex/bhex as indicated by
$0.27 \leq G \leq 0.31$. Buckling turns planar crossings into avoided crossings and can split a single $\Gamma$-loop into two concentric loops; this raises $N_{\mathrm{cross}}/L$ but keeps the near-$E_F$ states $s/p$-dominated in Cu/Ag (Figures S33–S34~\cite{SM}). Au shows the same tendencies, but with larger splittings, where the $d$-band branches approach $E_F$.

Zn and Cd have most of their Fermi contours around $\Gamma$ and exhibit additional loops and small pockets, compared to the coinage metals (Figures S36-S37~\cite{SM}). In planar hex, $A_{\mathrm{line}}$~is~$0.08$ and $G~\in [0.54, 0.59]$ indicate a nearly circular, $\Gamma$-centric loop~(Table~S1-S2). $F_{\mathrm{flat}}$ is low-to-moderate with Zn(hex) having $\sim$ 0.28, and Cd(hex) $=$ 0.15, in line with extended low-slope segments on a round loop. In sq, $F_{\mathrm{flat}}$ is larger Cd(sq) $\gtrsim$ 0.28, and the loop elongation grows. Buckling to bsq drives a strong anisotropy,
with $0.78 \leq A_{\mathrm{line}} \leq 0.82$ and can fragment the box-like loop into edge-centered “lenses” and corner “satellites”. Here, side loops mean small closed pockets which are detached from the main $\Gamma$-loop, and satellites are short secondary arcs that wrap around a larger loop; their appearance lifts $N_{\mathrm{cross}}/L$ markedly, with Zn(bsq) reaching 25. The presence of central pockets increases $G$ and reduces $A_{\mathrm{line}}$; corner/edge pockets keep $G$ smaller but still increase the $E_F$ crossings. Across both groups, $\langle v_F\rangle$ stays near $10^6\ \mathrm{ms^{-1}}$ with weak lattice dependence, while local $v_F$ maxima align with curved segments and pocket tips~(Figure~\ref{fig:vf_lattice_orbital}a).

Taken together,  coinage metals keep small $F_{\mathrm{flat}}$, large $A_{\mathrm{line}}$ in hex/sq and small $A_{\mathrm{line}}$ in bsq, moderate $G$ in hex/bhex, and high $N_{\mathrm{cross}}/L$ in hex/bhex. Group 12 shows higher $F_{\mathrm{flat}}$ in several lattices, the highest $N_{\mathrm{cross}}/L$ is in bsq, round central loops in planar hex (small $A_{\mathrm{line}}$, large $G$), and strong elongation under buckling. Thus, lattice type sets loop elongation and radius placement ($A_{\mathrm{line}}$ and $G$), buckling affects crossings and loop splitting ($N_{\mathrm{cross}}/L$), and exploit element-specific $s/p$$-d$ proximity to tune local flatness $F_{\mathrm{flat}}$ and Fermi velocity~(Figures.~\ref{fig:vf_lattice_orbital} and~\ref{fig:metrics_lattice_orbital}).

Among closed-$d$-shell metallenes, Hg is a special case. DFT-HSE calculations show the band gaps in all six lattices: 1.55 eV for bsq, 1.70 eV for bhc, 1.97 eV for sq, 2.47 eV for bhex, 2.49 eV for hex, and 3.55 eV for hc. Thus, corresponding Fermi-lines are absent altogether (Figure~ S38).

\subsection{$p$-band metallenes (Groups 13--15)}
\begin{figure*}[t!]
      \centering
    \includegraphics[width=0.99\linewidth]{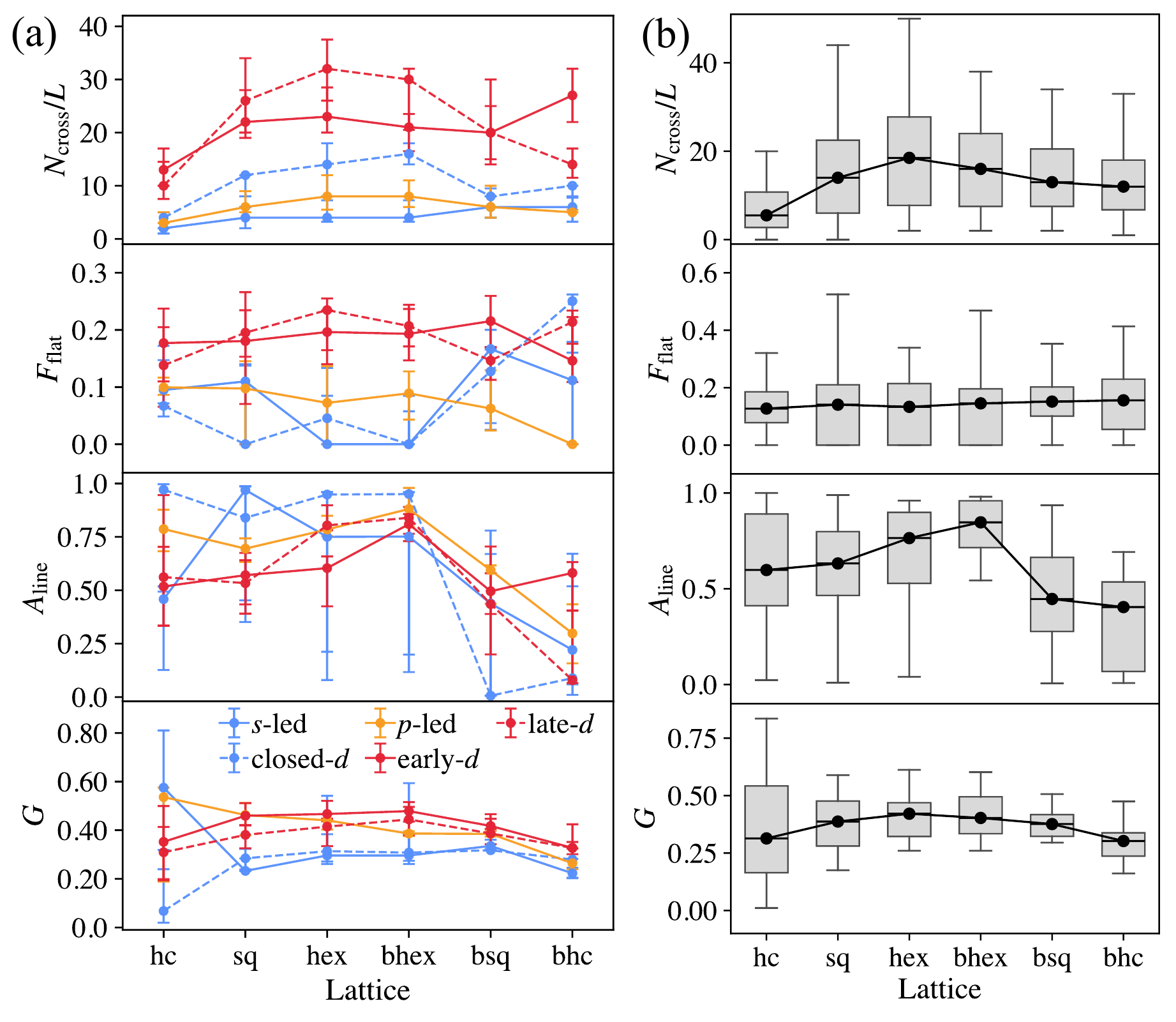}
    \caption{(a)~Orbital-resolved band structure and fermiology metrics vs lattice. For each of the six lattices, the panels show lattice-wise medians of $N_{\mathrm{cross}}/L$, $F_{\mathrm{flat}}$,  $A_{\mathrm{line}}$, and $G$ for $s$-led (Group~1-2), closed$d$~(Group~11-12), $p$-led~(Group~13-15), early-$d$-led~(Group~3-6), and late-$d$-led~(Group~7-10)  metallenes; error bars denote the interquartile range.~(b)~Lattice-resolved statistics of the four metrics. From top to bottom, boxplots show the distributions of $N_{\mathrm{cross}}/L$, $F_{\mathrm{flat}}$, $A_{\mathrm{line}}$, and $G$ for all elemental metallenes in the six lattices. For each lattice, the grey box and whiskers indicate the interquartile range and overall spread of the metric, while the black dot and connecting line mark the lattice-wise median, highlighting how coordination and buckling systematically reshape multiband connectivity, band flatness, Fermi-line anisotropy, and $\Gamma$-centricity.~ Together, the trends reveal how Fermi-level crossing density, band flatness, Fermi-line anisotropy, and $\Gamma$-centricity depend jointly on lattice geometry and dominant orbital character.}
    \label{fig:metrics_lattice_orbital} 
\end{figure*}

In Groups 13–15, the states near $E_F$ are mostly $p$-states. 
The $p$-metallenes have comparable $\langle v_F\rangle$ to the 
$s$-metallenes, but with stronger lattice-to-lattice variation, especially when the lattices buckle~(Figure~~\ref{fig:vf_lattice_orbital}c).
With buckling, $p_z$ mixes with in-plane orbitals, and band crossings that exist in flat layers turn into small avoided crossings. Al is the exception and has near-parabolic shaped $s$-bands in several lattices with only modest $p$-orbital mixing. The Fermi contours are loops around $\Gamma$ with a few element-lattice pairs having small detached pockets (Figure~ S39-S45).

In the hexagonal geometry, the loops remain around $\Gamma$, but their shape and position change by element. The Al(hex) shows moderate flatness with many crossings and an elongated loop at mid radius, quantitatively, $F_{\mathrm{flat}}~=~ 0.15$, $N_{\mathrm{cross}}/L=22$, $A_{\mathrm{line}}\approx 0.75$, and $G\approx 0.44$. Ga, In, and Tl also carry elongated loops and mid-radius placement \big($A_{\mathrm{line}}\approx 0.82$–$0.88$, $G\approx 0.43$–$0.45$, and only a few crossings). Sn, Pb, and Bi have moderate $G$ and sizable elongation and roughly $N_{\mathrm{cross}}/L = 10$. Buckling to bhex further enhances loop elongation in Ga, In, and Tl, pushing $A_{\mathrm{line}}$ to its maximal values with fewer crossings, and it maintains strongly elongated loops in Sn, Pb, and Bi with high $A_{\mathrm{line}}$ but a with a larger number of crossings. These values indicate that hex and bhex keep a stable $\Gamma$-centred loop. Buckling strengthens the sixfold shape and can create or remove small closed loops near the zone edges, which raises $N_{\mathrm{cross}}/L$~(Figure~\ref{fig:vf_lattice_orbital}b-c). $F_{\mathrm{flat}}$ stays low for most $p$-metallenes, indicating the steepness in bands, but it increases locally in Sn, Pb, and Bi near avoided crossings, which mark small low-slope segments. The appearance, growth, shrinking, or merging of these small closed loops away from $\Gamma$ can be termed as pocket rearrangement, which remains element-dependent.

In the square lattice, the Fermi-lines are more rounded and lie closer to the zone center and moderate elongation with $G\approx 0.46$–$0.53$ and \big[$A_{\mathrm{line}}\approx 0.52$–$0.70$ for Al, Ga, In, and Tl~(Figure~\ref{fig:metrics_lattice_orbital}). The bands possess limited flat segments with small $F_{\mathrm{flat}}$ small and are mostly dispersive with $N_{\mathrm{cross}}/L=4$–$12$\big].
Sn(sq), Pb(sq), and Bi(sq) lean toward stronger elongation or additional pockets \big(Sn: $A_{\mathrm{line}}\approx 0.78$, $G\approx 0.33$; Pb: $A_{\mathrm{line}}\approx 0.71$, $G\approx 0.37$; Bi: $A_{\mathrm{line}}\approx 0.96$, $G\approx 0.39$, $N_{\mathrm{cross}}/L=10$\big). Upon buckling, $F_{\mathrm{flat}}$ decreases for In, Sn, and Pb; and $N_{\mathrm{cross}}/L$ decreases for Ga, In, and Bi. Buckling makes the Fermi-lines more isotropic, except for Al, for which $A_{\mathrm{line}}$ increases, and these Fermi contours shift away from $\Gamma$ for Group 13 and 15 metallenes as $G$ decreases but move towards $\Gamma$-point for Sn and Pb. 

 In hc, Al to Tl have large centricity \big(Al: $G\approx 0.54$; Ga: $G\approx 0.56$; In: $G\approx 0.59$; Tl: $G\approx 0.60$\big) and high elongation \big($A_{\mathrm{line}}\approx 0.74$–$0.95$\big). Sn, Pb, and Bi shift the weight outward \big($G\approx 0.15$–$0.23$, with $A_{\mathrm{line}}\approx 0.60$–$0.79$\big). The most level bands are those of In(hc), followed by Tl(hc), and the steepest bands are those of Sn(hc). The crossing density is highest  for Al(hc) and lowest for Bi(hc), other metallenes. Buckling promotes a central channel for most elements. Al becomes nearly circular \big($A_{\mathrm{line}}\approx 0.10$, $G\approx 0.24$, $N_{\mathrm{cross}}/L=16$\big). Ga, In, and Tl keep moderate centricity \big($G\approx 0.23$–$0.28$\big) with $N_{\mathrm{cross}}/L=5$. Sn and Pb move inward \big($G\approx 0.34$–$0.35$, $N_{\mathrm{cross}}/L=5$–$6$\big). Bi(bhc) shows a contact-like case with very few crossings. These numbers demonstrate how breaking mirror symmetry introduces a central pocket or splits a loop, increases $N_{\mathrm{cross}}/L$ in certain cases, and alters elongation.

\subsection{Early transition $d$-metallenes (Groups 3--6)}
\begin{figure*}[t!]
      \centering
    \includegraphics[width=0.99\linewidth]{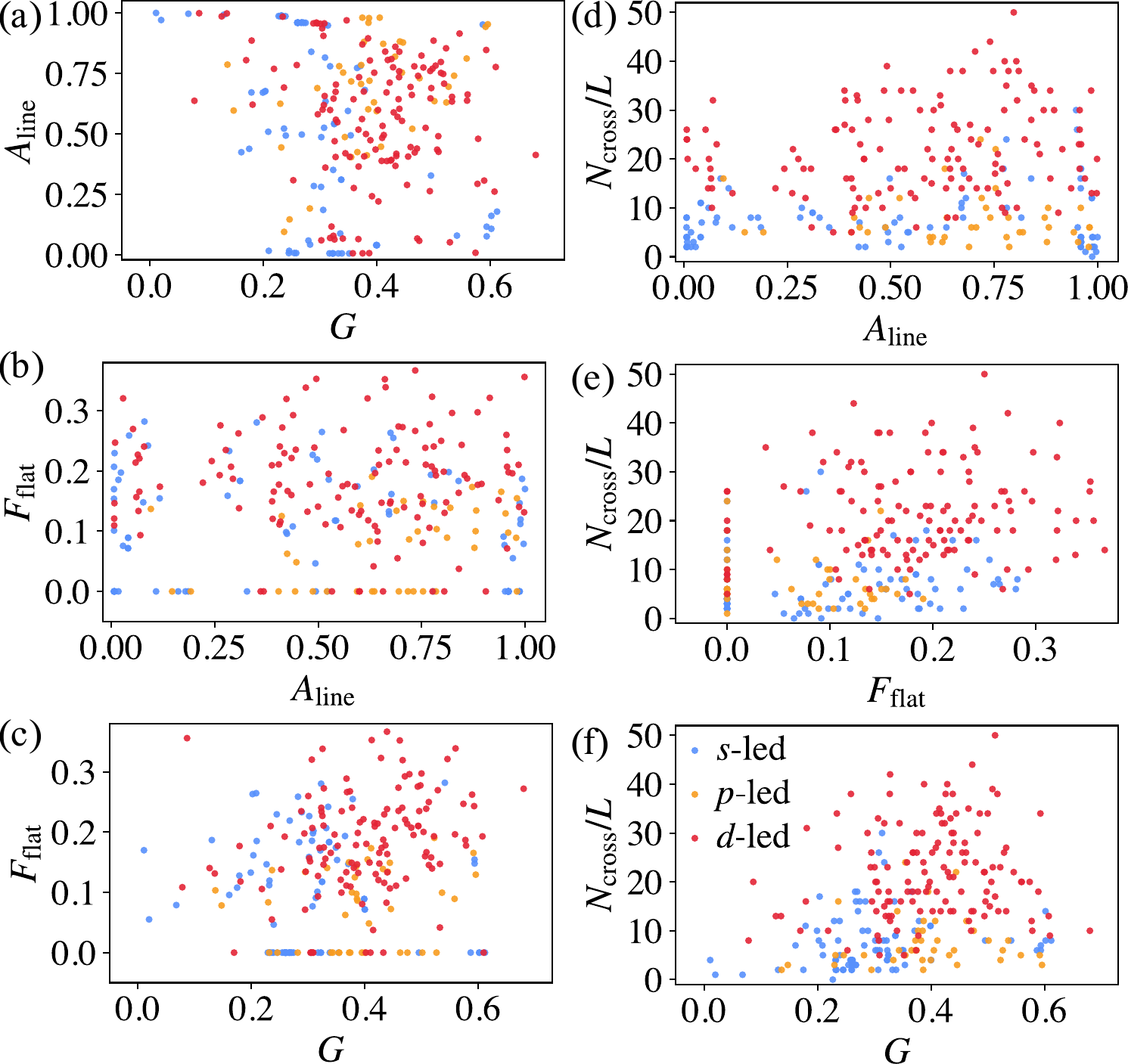}
    \caption{Correlation between metrics across all metallenes.~(a-f)~Scatter plots show the pairwise relationships between the four metrics $N_{\mathrm{cross}}/L$, $F_{\mathrm{flat}}$, $A_{\mathrm{line}}$, and $G$. Each point corresponds to a single metallene and is colored by dominant orbital character:
$s$-led (Group~1-2,~11-12) (blue), $p$-led~(Group~13-15) (orange), and $d$-led~(Group~3-10) (red). The plots reveal how multiband connectivity, band flatness, Fermi-line anisotropy, and $\Gamma$-centricity co-vary and how $s$-, $p$-, and $d$-dominated metallenes occupy distinct regions of the correlation space.}
    \label{fig:correlation} 
\end{figure*}

In early transition metallenes, the $d$-states primarily determine the electronic structure, while the $s/p$-states contribute minimally. The presence of $d$-states shifts the $\langle v_F\rangle$ distributions to lower values than those of the $s$-, $p$-, and closed-$d$ metallenes, indicating overall slower carriers across the different lattices~(Figure~\ref{fig:vf_lattice_orbital}d).
Averaged over the lattices, the Fermi-level crossing density increases from the 3$d$-to the 4$d$-series only in Group 3. In other groups, as orbitals extend, the average crossing density decreases. Notably, in the Group 5 4$d$-series, Nb metallene exhibits the highest average crossing density. This indicates that the 5$d$ series metallenes often reduce multiband connectivity and intersections along the high-symmetry path.
Coordination also affects crossing density as it changes from 3 (hc) to 4 (sq); the crossing density nearly doubles, and when it changes from 4 (sq) to 6 (hex), it has only a slight effect. 
Buckling also affects the crossing density and most often increases it. 
The buckling to hc lattice increases the crossings $\sim$ 5.2$\times$ for Sc, and for most metallenes $\sim$ 2$\times$ with the least increase observed for Cr ($\sim 1.3$$\times$). However, when sq lattice buckles, the situation is different, and the crossing density increases only for Ti, Hf, Nb, and W, with the highest increment for Hf ($\sim$~1.6$\times$). Also, buckling to a hex lattice can either reduce or lift crossings depending on whether long edge arcs are gapped or doubled (for example, Ti (26$\to$14) drop; Ta (20$\to$24) rise).

The band flatness $F_{\mathrm{flat}}$ is largest in sq, moderate in hex, and lowest in hc lattices. In a given Group, metallenes of the 4$d$-series generally exhibit the highest flatness score, $F_{\mathrm{flat}}$, compared to their 3$d$- and 5$d$- counterparts. For example: Ti $\to$ Zr $\to$ Hf shows a decrease; V $\to$ Nb shows a peak, then drops at Ta; and Cr $\to$ Mo $\to$ W decreases steadily, reflecting broader and steeper dispersions. If we consider, for example, Mo(sq) with $F_{\mathrm{flat}}~=~0.52$, it reflects long low-slope segments near the square-edge saddles. In hex, flatness is typically moderate: Hf(hex) has $F_{\mathrm{flat}}~=~0.32$ and Ta(hex) $\approx 0.20$, while Nb(hex) is lower at $\approx0.15$. The honeycomb family generally shows the least flatness unless a $K$-sector touching sits very close to $E_F$; otherwise, the steep $\Gamma$-centered $s$-band parabola and dispersive $d$-bands keep $F_{\mathrm{flat}}$ small.
Buckling modifies $F_{\mathrm{flat}}$ by turning band crossings into mini-gaps or by pushing van-Hove-adjacent states toward $E_F$. When gaps appear along the sampled BZ paths, $F_{\mathrm{flat}}$ rises: Ta(hex to bhex) increases from $\approx$ 0.20 to 0.30, and Mo(hex to bhex from $\approx$~0.08 to 0.24 as new avoided crossings create extended low-slope arcs. There are also counter-examples where buckling reduces flatness because dispersion steepens along the BZ path or a flat segment moves away from $E_F$; Zr(hex to bhex) drops from $\approx$~0.17 to 0.14. In the sq family, buckling preserves a high $F_{\mathrm{flat}}$ when the $C$ saddle remains near $E_F$, or lowers it if the states reorganize away from saddle-adjacent regions.

For the shape of the Fermi-lines, the measure $A_{\mathrm{line}}$ tends to be small for hc compared to sq and hex lattices, where Fermi-lines stretch along the edges and corners of the zone. For example, V(hc) has $A_{\mathrm{line}}\approx 0.03$ with a circular near $\Gamma$-loops with large $G\sim0.83$, however, for Zr(hc), the Fermi-line breaks into six edge-following arcs with no central loop, so $A_{\mathrm{line}}$ is large $\approx 0.89$ and $G$ is small. In hex, Mo and W are strongly elongated with $A_{\mathrm{line}}$ $\approx$0.75, 0.71, consistent with boundary follow contours.

Across Group 3, (Sc$\to$Y) often shows a decrease in $A_{\mathrm{line}}$ for several lattices, Y is more isotropic than Sc; this isotropy comes with a rise in $G$ that is, loops pull inward toward $\Gamma$. Buckling can partially reverse this by elongating edge–aligned paths and lowering $G$; for example, for Sc, hex to bhex $A_{\mathrm{line}}$ increases from $\approx$ 0.50 to $\approx$ 0.97. In Group 4 (Ti$\to$Zr$\to$Hf), anisotropy is strong for Ti/Zr in sq/hex and often softens at Hf; Zr(hc) is an exception, while Ti(sq) is rounded and then becomes elongated in bsq when buckled. Notably, the Hf(hc/bhc) bands show an electron-like dispersion along $\Gamma$--$M$ crossing $E_F$ and producing a $\Gamma$-centered pocket (Figure S15~\cite{SM}), consistent with the Hf-derived ARPES band reported for hafnene on Ir(111)~\cite{Xiao2021_Hf_on_Ir(111)}. In the supported system, however, most other Hf-derived bands are largely quenched by strong substrate hybridization~\cite{Xiao2021_Hf_on_Ir(111)}.

In Group 5 (V$\to$Nb$\to$Ta), $A_{\mathrm{line}}$ typically peaks around Nb and drops at Ta; V(hc) is near-perfect isotropic with large $G$, whereas Nb(hex) remains elongated and more boundary–weighted (smaller $G$). In Group 6 (Cr$\to$Mo$\to$W), $A_{\mathrm{line}}$ tends to decrease down the series: Mo and W in hex stay elongated (e.g., $A_{\mathrm{line}}\approx0.75 ~\mathrm{and}~0.71$) but are more uniform than Cr in several lattices, and buckling can either accentuate elongation along edge paths or round a central pocket, raising $G$ (Figure~\ref{fig:correlation}a).
Overall, it is observed that when $A_{\mathrm{line}}$ is high in sq/hex,$F_{\mathrm{flat}}$ is often high too (Figure~\ref{fig:correlation}b).
The $N_{\mathrm{cross}}/L$ tends to increase when $A_{\mathrm{line}}$ rises large, as the added anisotropy comes from multiple bands at $E_F$. Low $A_{\mathrm{line}}$ usually accompanies a more $\Gamma$-centric loop (higher $G$) and fewer crossings~(Figure~\ref{fig:correlation}c, d).

\subsection{Late transition $d$-metallenes (Groups 7--10)}
We now turn to the highest filling $d$-state manifolds, spanning the half-filled row (Group~7) through the nearly filled row (Group~10).
These metallenes span the widest range for $\langle v_F\rangle$ distributions, with some lattices hosting slower states than the early 
$d$-metallenes, while others reach median comparable to or exceeding those of the $s$-metallenes~(Figure~\ref{fig:vf_lattice_orbital}e).~All metallenes possess a $d$-led near-$E_F$ spectrum, while $\Gamma$-centered $s$- and $p$-states participation remains lattice- and element-dependent.
In hc, 3$d$ series metallenes have large $N_{\mathrm{cross}}/L$ and moderate flatness $F_{\mathrm{flat}}$: Mn(hc)/Fe(hc) has $F_{\mathrm{flat}}\approx$ 0.21/0.36 with $N_{\mathrm{cross}}/L=18/20$. Co(hc) and Ni(hc) have higher crossing density but are less flat. In contrast, 4$d$ and 5$d$ metallenes have steep bands with low crossing density. In general, sq lattice amplifies $N_{\mathrm{cross}}/L$; for example, Mn/Fe in sq now has $N_{\mathrm{cross}}/L=42/38$. However, in sq lattices, bands are less steep $F_{\mathrm{flat}}$(sq)~$>$~$F_{\mathrm{flat}}$(hc), except for Fe, Ir and Pd. The hexagonal lattices have the crossings and flatness along the path $X-C-Y$. The crossing density is maximal for late row metallenes: the $N_{\mathrm{cross}}/L$ = 50 for Os(hex) with $F_{\mathrm{flat}}$ =~0.25, and it is 40 for Ir(hex) with $F_{\mathrm{flat}}$~= 0.32, Pt is more moderate with crossing density of 32 and flatness 0.13. The Ni/Pd are having less dense bands near $E_F$ with $N_{\mathrm{cross}}/L$ = 23/15 and $F_{\mathrm{flat}}$~=~0.26/0.20. The Co/Rh carry many crossings, 26/35, with less flatness, 0.07/0.04. 

Buckling shifts crossing density and flatness in lattice-specific ways. In hc to bhc, Os, Ni, and Pt gain flatness and crossings, while Fe, Pd, and Ir lose band flatness but gain more crossings. When sq buckles to bsq, buckling often redistributes electronic states, and orbital mixing opens small gaps where bands used to cross. Long, flat parts of the bands or Fermi-lines (slow electrons, heavy mass) get broken up~(Figure~\ref{fig:vf_lattice_orbital}e). This was inferred by Pt and Co, for example. However, a few metallenes (Rh and Ir) gain flatness, and as the avoided crossings open, the crossing density goes down.  In hex, buckling tends to trade crossings for flatter segments on the edge rails for many late rows metallenes like Os, Re, and Rh. Others lose flatness and slightly adjust crossings, for instance, the $F_\mathrm{flat}$~changes from 0.32 to 0.16 for Ir and $N_{\mathrm{cross}}/L$ from 40 to 38~(Figure~\ref{fig:bands_fermi_schematics}d). For Pt, the change for $F_\mathrm{flat}$ is from 0.13 to 0.11, and crossings increase ~(from 32 to 34). Mn is a strong outlier: flatness jumps from 0.20 to 0.47 while crossings fall from 40 to 30.

The Fermi-line segments in hc lattices split into two patterns. Some metallenes keep central structure but with very different anisotropy–centricity: Mn(hc) is moderately anisotropic with $A_{\mathrm{line}}\approx0.29$ and mid-centricity $G\approx0.45$; Fe(hc) becomes almost maximally anisotropic with small $G$; Re(hc) sits between them. Co/Ni(hc) lack a near $\Gamma$ loop and are strongly perimeter-weighted with low $G\approx0.14/0.13$. Buckling drives these toward more centered, less anisotropic sheets: Fe/Ru/Os/Co/Rh/Ni(bhc) becomes nearly isotropic. In sq lattices, Fe/Ru/Os show the nested-core plus corner-oval geometry with increasing centricity from Fe to Os: Fe(sq) has high anisotropy and low centricity $(A_{\mathrm{line}}\approx0.78,\ G\approx0.26)$; Ru(sq) moderates (0.60,\ 0.40); Os(sq) is more centered (0.37,\ 0.43). Pd(sq) is the most isotropic $\Gamma$-centered dominated case $(A_{\mathrm{line}}\approx0.01,\ G\approx0.57)$, whereas Co(sq) stays highly anisotropic and edge/perimeter-led (0.98,\ 0.23); Ir/Pt(sq) lie between (0.53,\ 0.41) and (0.65,\ 0.32). 

Buckling splits the inner Fermi-line segments and shifts spectral weight: Fe(bsq) and Ni(bsq) become near-circular, Rh(bsq) retains strong anisotropy, and Pt(bsq), Pd(bsq), Ir(bsq) can be referred to as in the mid-anisotropy band. In hex lattices, Pd/Rh/Ir/Os/Pt share an hourglass-like edge with bright vertical flanks and moderate centricity: Pd(hex) ($A_{\mathrm{line}}\approx0.78,\ G\approx0.50$), Rh(hex) (0.84,\ 0.41), Ir(hex) (0.78,\ 0.43), Os(hex) (0.80,\ 0.51), Pt(hex) (0.81,\ 0.42). Fe/Co/Ni(hex) are the cleanest quasi-1D edge-parallel segments with very high anisotropy and low $G$. Buckling keeps these segments as the main feature, but subtly recenters the 4$d$/5$d$ series metallenes and shifts the Fermi-line modestly toward $\Gamma$~(Figure~\ref{fig:correlation}a-f).

\section{Discussion}

It is found that the coordination controls the Fermi-line loop shape and their placement, while buckling redistributes band crossings and creates or merges pockets~(Figures.~\ref{fig:bands_fermi_schematics}~and~\ref{fig:vf_lattice_orbital}). To quantify all these effects we used the $N_\mathrm{cross}/L$, $F_\mathrm{flat}$, $A_\mathrm{line}$, and $G$. These lattice- and orbital-averaged trends are summarized in Figure~\ref{fig:vf_lattice_orbital}.

Lattice medians show that crossing density is lowest for hc and highest for hex. Buckling lowers the crossing density; the bhex and bsq lie between sq/hex and hc, and bhc has a larger median $N_{\mathrm{cross}}/L$ than hc but remains below those of sq and hex. These medians come with broad spreads, measured by the median absolute deviation (MAD), and are highest for sq and hex, in the range of $8-11$. This indicates substantial element-to-element variation in crossing density across these lattices.
Flatness varies more gently, but still follows a systematic trend. Planar lattices have a lower median $F_{\mathrm{flat}}$, and buckled lattices shift it upward by a small amount. These shifts are small compared with the intra-lattice spread, whose MAD is of order $0.05-0.10$. This indicates that lattice choice modulates flatness only weakly compared with element-to-element variation~( Figure~\ref{fig:vf_lattice_orbital}b and Tables S1–S2).

The anisotropies of the Fermi-lines, measured by $A_\mathrm{line}$, show that hex and bhex have the most elongated contours, with the highest medians among all lattices. The sq and hc are moderately elongated, and bsq and bhc are most isotropic with the roundest contours. The MAD of $A_\mathrm{line}$ is of the order $0.18-0.27$ for hc, sq, hex, bsq and bhc, but it is 0.11 for bhex. This lower MAD in bhex means that the values of $A_\mathrm{line}$ cluster more tightly around a large median. The $\Gamma$-centricity $G$ follows these trends. The median of $G$ is highest for hex/bhex, followed by sq/bsq, and lowest for hc/bhc. This means that hex and bhex place the large fraction of their Fermi-line near $\Gamma$-point, whereas hc and bhc lattices distribute the Fermi-line contours more broadly across the Brillouin zone. For all lattices except hc, the MAD of $G$ is below 0.10; for hc, it is around 0.17. This spread is small compared to the differences between lattice medians, so the ranking $G({\text{hex/bhex}}) > G({\text{sq/bsq})} > G({\text{hc/bhc}})$ remains essentially unchanged when the metallene type changes within each lattice.

These trends give a coherent picture of how bands and Fermi-line geometry respond to lattice and element choice. Changing the lattice, especially introducing buckling, shifts $F_\mathrm{flat}$  only by a small absolute amount compared with the variation between elements. Still, it systematically reduces $A_\mathrm{line}$ and therefore shortens long straight segments into more locally curved contours. At the same time, $N_{\mathrm{cross}}/L$ and $G$ are much more lattice sensitive. The sq and hex lattices support the highest and most variable crossing densities and a strong concentration of Fermi-line length around $\Gamma$, while buckled lattices pull these trends toward intermediate values. Element type then moves each metallene within this lattice-defined framework, but the overall ranking in crossing density, elongation, and $\Gamma$-centricity is set primarily by the lattice.

To combine the behavior of bands and Fermi-lines we introduce a new quantity, an element-level pocketness score $\mathcal{P}$. $\mathcal{P}$ compresses four descriptors: band flatness $F_{\mathrm{flat}}$, Fermi-line anisotropy $A_{\mathrm{line}}$, $\Gamma$-centricity $G$, and $N_{\mathrm{cross}}/L$  into a single number. We first orient these descriptors so that “pocket-like” behavior always corresponds to features: slow bands with sizeable flat segments, round rather than elongated contours, Fermi-lines concentrated near $\Gamma$, and fewer Fermi-level crossings. A simple data-fitted weighting then combines these four oriented ingredients into a single scalar~$\mathcal{P}$~(details are in SI). 
High $\mathcal{P}$ marks elements whose Fermi-lines look like compact pockets, isotropic and closer to $\Gamma$, and the bands that are slow over finite $k$-intervals with only a few crossings of $E_F$. Low $\mathcal{P}$ signals the opposite, where bands cross $E_F$ frequently and Fermi-lines form extended, strongly elongated contours that thread across the Brillouin zone and carry more weight toward its edges.
\begin{figure}[t!]
      \centering
    \includegraphics[width=0.99\linewidth]{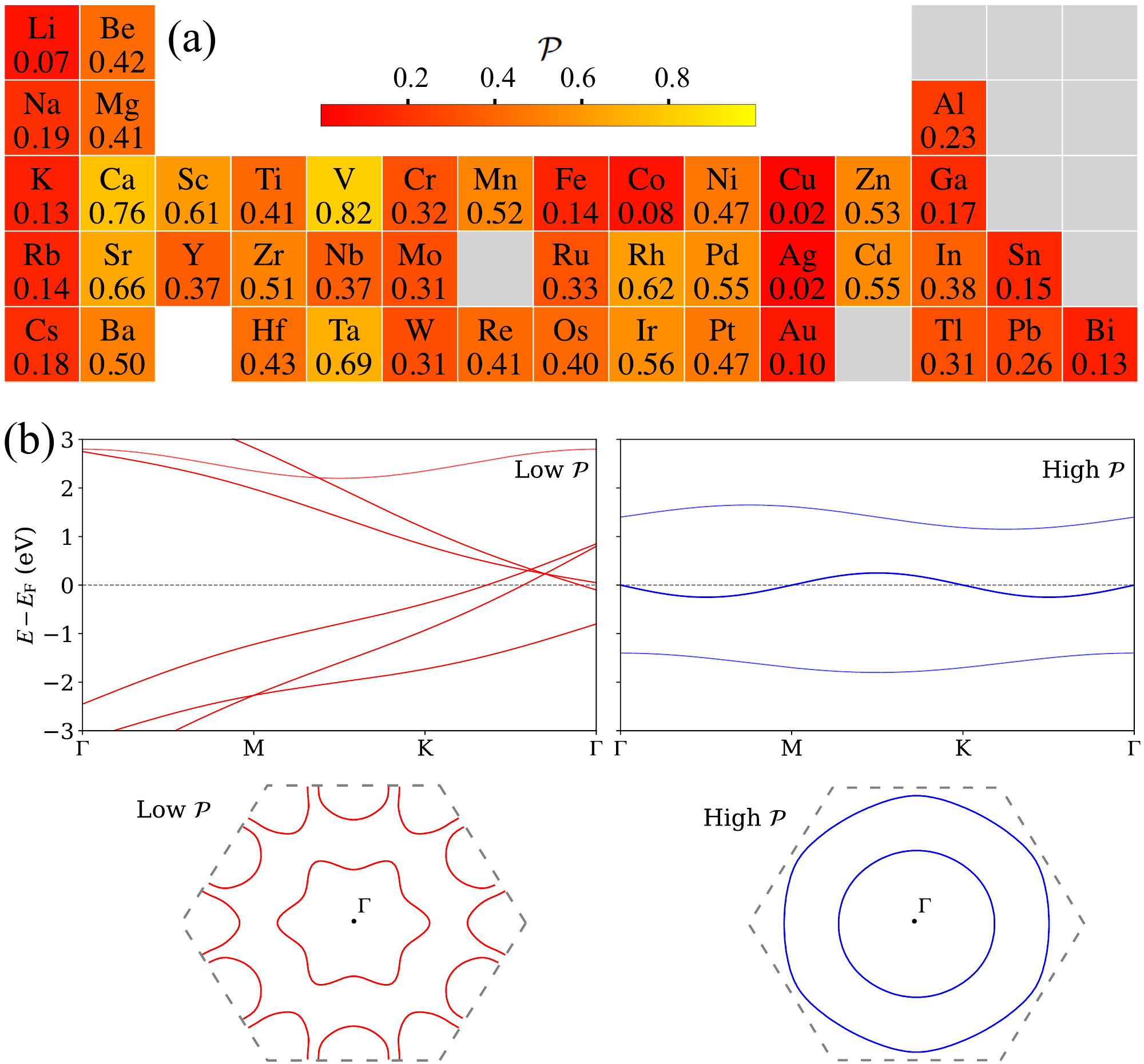}
       \caption{(a)~Identifying metallenes on the basis of $\mathcal{P}$, a metrics-weighted per-element score for locally flat bands with fewer $E_F$ crossings and nearly isotropic Fermi contours which sits near $\Gamma$ (details in SI). Low-$\mathcal{P}$ values cluster among the alkali and coinage metallenes together with a few light $p$-led and 3$d$ metallenes, the majority of the remaining main-group and transition metallenes have a moderate value of $\mathcal{P}$, whereas high-$\mathcal{P}$ are of alkaline-earth and early/late $d$-metallenes.~(b)~Schematic band structures and Fermi-line topologies representative of low- and high-$\mathcal{P}$ behavior. Low $\mathcal{P}$ shows many steep crossings of $E_F$ and Fermi-line contours whose radius changes a lot and remain close to the FBZ edges, while high $\mathcal{P}$ shows fewer, flatter bands near $E_F$ and compact, nearly isotropic $\Gamma$-centred loops. }
    \label{fig:pocketness} 
\end{figure}
This pocketness score is, to our knowledge, the first element-resolved scalar descriptor built from the band and Fermi-line descriptors. It allows us to map how band structure and Fermi-line character vary across the periodic table (Figure~\ref{fig:pocketness}). It is also worth emphasizing that we confirmed the robustness of $\mathcal{P}$ with respect to reasonable choices made in its construction by conducting sensitivity analyses, and that its behavior is not dominated by arbitrary scaling decisions or the tuning of any single parameter.~(See Supplemental Material~\cite{SM})

The resulting ordering is chemically reasonable: highest $\mathcal{P}$ concentrate in early $d$ and alkaline-earth metallenes (for example: V, Ta, Sr, Sc), while coinage and some light $s$-metallenes have low $\mathcal{P}$~(Figure~\ref{fig:pocketness}). These extrema correspond to the qualitative trend, for example, of coinage metallenes which keep elongated outer $\Gamma$-loops on hex/sq; Zn/Cd gain pockets under bsq; and many late-row $d$-metallenes exchange crossings for flatter edge-parallel when buckled.  In brief, geometry sets elongation and centricity, buckling toggles crossings and pocket activity, and $\mathcal{P}$ turns the four metrics into a single measure.

Taken together, the four metrics obtained from bandstructure and Fermi-lines with the per-element pocketness score give a compact, quantitative map from lattice geometry and orbital makeup to Fermi-surface shape and transport-relevant features~(Figures.~\ref{fig:vf_lattice_orbital},~\ref{fig:metrics_lattice_orbital},~\ref{fig:correlation}~\ref{fig:pocketness}). The quantitative investigation leads to design rules. To obtain round, central pockets with light effective masses, prefer lattices and metallenes which have $A_{\mathrm{line}}$ toward the range $\sim 0.40-0.45$ while maintaining $G$ in the range $\sim 0.40-0.42$; moderate buckling can help by localizing curvature and splitting long edge-parallel segments without driving the contours off-center. For isotropic in-plane transport and stable Shubnikov–de Haas~(SdH) oscillations, target high $\mathcal{P}$ metallenes in the early $d$ or alkaline-earth families; for anisotropic responses and large angle-dependent magnetoresistance, the hex/bhex geometries with higher $A_{\mathrm{line}}$ is preferable. The small absolute differences in $F_{\mathrm{flat}}$ medians across lattices mean flatness should be interpreted locally at the pocket perimeter rather than globally across the path, which explains why modest increases in $F_{\mathrm{flat}}$ can still coincide with strong pocketness when $A_{\mathrm{line}}$ is low and $G$ is high. The median and MAD analysis provides additional insights for each lattice. This analysis comes with caveats: metrics are extracted from high-symmetry paths and constant-energy contours at $E_F$, so ultra-fine features off the path are not explicitly resolved, but the overall trends remain consistent and applicable. The immediate implication is a screening workflow: filter by high $\mathcal{P}$, then verify desired $A_{\mathrm{line}}$ and $G$ on the target lattice under intended buckling or strain. This turns fermiology into a practical prescription for substrate selection, strain/buckling engineering, and element choice in pocket-driven device concepts.

\section*{Summary and Conclusions}

This study gives a compact, usable picture of how lattice and element choice shape the Fermi-lines in 270 monolayers of elemental metallenes. It was found that lattice sets the basic shape and placement of the Fermi-line contours, while out-of-plane buckling shortens long straight segments, and can create or merge small pockets by slightly raising the local flatness. The single pocketness score was used to capture these effects for each element. Early $d$- and alkaline-earth elements tend to form round, central pockets, while coinage and some light $s$ elements tend to be elongated and away from the BZ-center.

The proposed pocketness score, $\mathcal{P}$, consolidates the four metrics into a single, easy-to-use guide for selecting systems for specific purposes and for further investigations.
For example, large $\mathcal{P}$ metallenes, when subjected to sufficiently strong magnetic fields, will become SdH-favourable~\cite{bangura2008small}. Also, the metallenes with direction-dependent in-plane transport properties are those having a lower value of $\mathcal{P}$. 
Furthermore, experimental Fermi-surface mapping by ARPES is more straightforward~\cite{sobota2021angle,kordyuk2014arpes} in cases where a single closed pocket produces a bright, nearly circular ring near $\Gamma$ with low intensity from other bands. In contrast, multiple pockets or elongated, off-centre contours add extra arcs that overlap in momentum space, complicating the mapping~\cite{sobota2021angle,kordyuk2014arpes,borisenko2022fermi}. Higher $\mathcal{P}$ points to the first case. Thus, $\mathcal{P}$ provides a concise and reasonable pathway for shortlisting of metallenes for SdH studies, angle-dependent transport, and clean band-mapping experiments.

Our reference point is freestanding monolayers at hybrid-functional DFT; substrates, moiré potentials, disorder, and many-body corrections~(GW/DMFT) will shift numbers but not trends. Coupling these band-contour descriptors to substrate phase diagrams and folding them into high-throughput screens can guide the development of device concepts in catalysis, plasmonics, and orbitronics.
 
\section*{Acknowledgments}
We acknowledge the Vilho, Yrjö, and Kalle Väisälä Foundation of the Finnish Academy of Science and Letters and the Jane and Aatos Erkko Foundation for funding (project EcoMet) and the Finnish Grid and Cloud Infrastructure (FGCI) and CSC—IT Center for Science for computational resources.
%

\end{document}